%% file: main.tex
\documentclass[10pt,twocolumn]{IEEEtran}
\usepackage{multicol}

\usepackage{color}
\usepackage{soul}
\usepackage{amsmath}
\usepackage{graphics}
\usepackage{setspace}
\usepackage{cite}
\usepackage{latexsym}
\usepackage{float}
\usepackage{epsfig}
\usepackage{multirow}
\usepackage[table,xcdraw]{xcolor}
\usepackage{cite,cases,url}
\usepackage{amssymb}
\usepackage{graphicx}
\usepackage{epstopdf}
\usepackage{balance}
\usepackage{subcaption}
\usepackage{enumerate}
\usepackage{array}
\usepackage{algorithmic}
\usepackage{algorithm}
\usepackage{enumitem}
\usepackage[normalem]{ulem}
\usepackage{longtable}
\usepackage{ragged2e}
\usepackage{booktabs}
\usepackage{tabularx}

\useunder{\uline}{\ul}{}

\newcolumntype{L}{>{\centering\arraybackslash}m{5cm}}
\newcolumntype{K}{>{\centering\arraybackslash}m{6cm}}
\newcolumntype{P}{>{\centering\arraybackslash}m{2.3cm}}
\newcolumntype{M}{>{\raggedright\arraybackslash}m{2cm}}
\newcolumntype{N}{>{\raggedright\arraybackslash}m{2.5cm}}

\begin{document}

\title{
Prototyping O-RAN Enabled UAV Experimentation for the AERPAW Testbed}
\author{
\IEEEauthorblockN{Joshua Moore, Aly S. Abdalla, Charles Ueltschey and Vuk Marojevic \\
} 
\normalsize\IEEEauthorblockA{Department of Electrical and Computer Engineering,  Mississippi State University, USA} \\
\normalsize\IEEEauthorblockA{\{jjm702,asa298,cmu32,vm602\}@msstate.edu}
}
\maketitle
\begin{abstract}
\input{include/abs.tex}
\end{abstract}
\IEEEpeerreviewmaketitle

\section{Introduction}
\label{sec:intro}
\input{include/intro.tex}



\section{
O-RAN Enabled UAV Research Platform}
\label{sec:background}

\input{include/background}




\section{
\textcolor{black}{Experimental Data Collection, Generation, and Analysis
}}
\label{sec:data}
\input{include/data.tex}

\section{{\textcolor{black}{O-RAN Integration Challenges and R\&D Opportunities}} 
}
\label{sec:challenges}
\input{include/challenges.tex}




\section{Conclusions}
\label{sec:conclusions}
\input{include/conclusions.tex}

\section*{\textcolor{black}{Acknowledgment}}
\footnotesize This work was partially supported by NSF awards 2120442 and 1939334, and by the Office of Naval Research under Award No. N00014-23-1-2808.

\balance

\bibliographystyle{IEEEtran}
\bibliography{aly}

\section*{Biographies}
\footnotesize
\vspace{0.2cm}
\noindent
\textbf{Joshua Moore} (jjm702@msstate.edu)
is a PhD student in the Department of Electrical and Computer Engineering at Mississippi State University, Starkville, MS, USA. His research interests include O-RAN, 5G/next-G communications, and RAN Management and Orchestration.

\vspace{0.2cm}
\noindent
\textbf{Aly Sabri Abdalla} (asa298@msstate.edu)
is an Assistant Research Professor in the Department of Electrical and Computer Engineering at Mississippi State University, Starkville, MS, USA. His research interests are on wireless communication and networking, software radio, spectrum sharing, wireless testbeds and testing, and wireless security with application to mission-critical communications, open radio access network (O-RAN), unmanned aerial vehicles (UAVs), and reconfigurable intelligent surfaces (RISs).

\vspace{0.2cm}
\noindent
\textbf{Charles Ueltschey}
(cmu32@msstate.edu) is an undergraduate student in computer science at Mississippi State University, Starkville MS, USA. His research interests are Wireless communications, O-RAN security, and Machine Learning.

\vspace{0.2cm}
\noindent
\textbf{Vuk Marojevic} (vuk.marojevic@msstate.edu) is a professor in electrical and computer engineering at Mississippi State University, Starkville, MS, USA. His research interests include resource management, vehicle-to-everything communications and wireless security with application to cellular communications, mission-critical networks, and unmanned aircraft systems.

\end{document}

%% file: include/abs.tex
The Open Radio Access Network (O-RAN) architecture is reshaping the telecommunications landscape by enhancing network flexibility, openness, and intelligence. 
This paper establishes the requirements, evaluates the design tradeoffs, and introduces a scalable architecture and prototype 
of an open-source O-RAN experimentation platform within the \textcolor{black}{Aerial Experimentation and Research Platform for Advanced Wireless (AERPAW)}, an at scale testbed that integrates unmanned aerial vehicles (UAVs) with advanced wireless network technologies, offering experimentation in both outdoor testbed and emulation via a custom \textcolor{black}{digital twin (DT)}. Through a series of aerial experiments, we evaluate FlexRIC, 
an open-source RAN Intelligent Controller, within the AERPAW hardware-software platform for network data monitoring, providing valuable insights into the proposed integration and revealing opportunities for leveraging 
O-RAN to create custom service based optimizations for cellular connected UAVs. We discuss the challenges and potential use cases of this integration and demonstrate the use of a generative artificial intelligence model for generating realistic data based on collected real-world data to support AERPAW's \textcolor{black}{DT}.

%% file: include/intro.tex

The advent of the Open Radio Access Network (O-RAN) architecture marked a significant advancement in telecommunications, enhancing flexibility, openness, and intelligence across diverse network configurations. O-RAN’s modular design, featuring open interfaces, supports real-time resource management and artificial intelligence (AI) driven optimizations, fostering interoperability and innovation. This allows operators to tailor networks to evolving demands, reduce operational costs, and enhance service delivery. Key components of the O-RAN architecture include the \textcolor{black}{near-real time} RAN Intelligent Controller \textcolor{black}{(near-RT RIC)} and the \textcolor{black}{non-RT RIC}. These components, O-RAN's disaggregated RAN, and the open interfaces enhance network flexibility, reduce costs via competitive vendor markets, and improve network intelligence~\cite{10024837}.

There are several efforts in the United States and the rest of the world to promote open RANs and O-RAN. One of the primary goals is vendor diversity, which spurs domestic deployment, allows customized network scaling with more control over cellular network technology and parameters, and supports technology adoption and evolution toward 6G in a given region or country. 

The Platforms for Advanced Wireless Research (PAWR) program in the United States plays a pivotal role in integrating O-RAN technologies into its diverse testbeds, providing critical infrastructure for exploring the practical applications of O-RAN in real-world settings. The PAWR program supports five Wireless Community Testbeds (WCTs) designed to provide a controlled environment where researchers and developers can experiment with and evaluate new wireless technologies, protocols, and applications. The \textcolor{black}{Cloud Enhanced Open Software Defined Mobile Wireless Testbed for City-Scale Deployment (COSMOS)} focuses on millimeter wave technology, providing an urban environment to explore advanced communications and collect data. The \textcolor{black}{Platform for Open Wireless Data-Driven Experimental Research (POWDER)} centers around sub-6 GHz deployments, offering a flexible, programmable environment for wireless experimentation and data collection. 
 \textcolor{black}{The Aerial Experimentation and Research Platform for Advanced Wireless (AERPAW)} integrates wireless communications with aerial platforms, exploring protocols for aerial networks and gathering relevant data. \textcolor{black}{The Agriculture and Rural Communities (ARA)} testbed focuses on providing connectivity solutions for rural and underserved areas, supporting a range of wireless technologies for exploring innovative approaches for broader network coverage. Colosseum is a large-scale wireless network emulator that offers a highly configurable environment for testing and evaluating advanced wireless technologies. These platforms feature software-defined radio (SDR) hardware and software along with specialized radio equipment.


In the realm of wireless research, the O-RAN architecture is reshaping how WCTs are used for experimentation and data collection as exemplified by O-RAN use cases in POWDER such as RAN slicing~\cite{Johnson22NexRAN}. At this time COSMOS, POWDER, and Colosseum offer O-RAN experiments while AERPAW and ARA 
consider offering them in the future~\cite{ara}. 
The existing O-RAN capabilities available through WCTs are either for an emulated \textcolor{black}{radio frequency (RF)} channel or with no or limited 
mobility~\cite{mushi2023open}. 
AERPAW offers distinctive advanced wireless research features, facilitating controlled \textcolor{black}{three-dimensional (3D)} mobility in an at-scale outdoor testbed and in a digital twin (DT).
While~\cite{mushi2023open} 
provides insights into some of the essential requirements for O-RAN experimentation within AERPAW, it does not actualize the integration of O-RAN into the testbed. 

\textcolor{black}{Unmanned aerial vehicles (UAVs)} require adaptive networks, which O-RAN facilitates, for reliable connectivity. 
This paper outlines the design requirements and evaluates the available deployment choices before proposing an architecture and prototype 
for constructing a reproducible open-source O-RAN experimentation platform for advanced wireless UAV research. This platform furnishes the necessary interfaces and performance metrics to facilitate extensive research opportunities involving cellular network-connected UAVs. Our proposed platform lays the groundwork and the first practical integration for pioneering O-RAN experiments in AERPAW, enabling research on AI-driven UAV communication, network, and trajectory optimization~\cite{alliance}. The contributions of this work are:

\begin{itemize}[left=0pt, labelsep=1em, itemsep=0.25em, topsep=0.25em, parsep=0pt, partopsep=0pt]
    \item {
    We describe the tradeoffs and design choices for 
    a testbed that combines O-RAN components with AERPAW’s capabilities, creating an open-source, experimental research platform for UAV communications (Section II). 
    }
    \item {
    We offer 
    experimental 5G key performance indicators 
    collected between a UAV and ground nodes 
    and introduce a generative AI (GAI) method \textcolor{black}{to generate new data samples based on collected data for AERPAW's DT}, enabling further research and analysis (Section III).}
    \item {
    We identify the key challenges in integrating 
    O-RAN research capabilities into AERPAW and propose practical solutions and future research and development (R\&D) opportunities for O-RAN enabled UAV communications (Section IV).}
\end{itemize}


%% file: include/background.tex
\input{Tables/ReqTable}

The integration of UAVs into cellular networks necessitates adaptable research platforms to match rapid technological advances. 
While the non-RT RIC is a logical extension of cellular industry's orchestration frameworks, the near-RT RIC and the E2 interface are the salient O-RAN features of most interest to researchers~\cite{upadhyaya2022prototyping} and are the focus of this paper.

\subsection{Requirements}

The capabilities necessary to facilitate an O-RAN enabled UAV research platform are decribed in continuation. 
\begin{itemize}
    \item \textbf{Dynamic Network Architecture:} The architecture must possess scalability, heterogeneity, and adaptability to accommodate rapid changes in deployment strategies and varying user demands. 
    It should be designed 
    to 
    facilitate the seamless integration of emerging technologies 
    and exhibit the flexibility to adapt to different use cases ranging from low altitude platforms, such as drones and tethered UAVs, to high altitude platforms. 
    \item \textbf{Standard Compliance:} Ensures 
    adherence to the latest 3GPP and O-RAN Alliance specifications for global interoperability and access to state-of-the-art technology. Compliance to 3GPP and O-RAN standards facilitates the integration of advanced technologies including network slicing, edge computing, and multiple-input multiple-output (MIMO) while enabling diverse user services. 
    \item \textbf{Open Interfaces and Modular Design:} 
    By integrating O-RAN’s open architecture, 
    the testbed design must be capable of offering a wide range of customization, modularity, and flexibility. \textcolor{black}{For instance, the testbed should facilitate the integration of diverse terrestrial or non-terrestrial radio unit (RU), distributed unit (DU), and central unit (CU) deployments over a wide range of vendors.} 
    Additionally, incorporating open interfaces, such as the E2 interface, 
    enables seamless communication between network components. 
    This flexibility and modularity enable efficient communication and coordination between various components of the O-RAN system, facilitating dynamic optimization and resource allocation in response to changing network conditions and traffic demands across different use cases and experimental requirements.  
    \item \textbf{AI-Enhanced RICs:} The architecture should support AI controllers with varying time-scale inference mechanisms for enabling research on 
    advanced network management. This includes non-RT inference for historical channel quality predictions, near-real-time inference for adjusting aerial node trajectories and transmission power, and real-time inference for resource block scheduling. By accommodating these processes, the architecture facilitates efficient decision-making from long-term planning to immediate network optimization. The design must consider the limited computational and power resources, particularly at aerial nodes, necessitating lightweight and energy-efficient algorithms to minimize overhead and power consumption.
    \item \textbf{Extensive Open-Source Software Support:} The O-RAN testbed should support a comprehensive suite of open-source software tools covering all network layers, simplifying the processes of modification, upgrading, and maintenance. Different software versions may needs to be made available for different experimenter hardware choices. 
    \item \textbf{Advanced Security Protocols:}  Robust security measures must be implemented to protect the network and its data from unauthorized access and cyber threats, safeguarding the integrity and confidentiality of research as well as experimental network and vehicle 
    operations.
    \item \textbf{Hardware and Software Codesign and Maintenance:} WCTs employ commercial off-the-shelf (COTS) hardware (computers, networking devices, software radios, RF components), which need to be maintained along with the software. COTS hardware balance performance with usability and reproducibility, ensuring programmability and cost-effectiveness in the deployment of network components.
    \item \textbf{\textcolor{black}{Open Application Programming Interfaces (APIs)} and Access to Software Development Tools:} The O-RAN testbed to be most valuable for researchers should offer unrestricted access to APIs and a comprehensive software toolkit, facilitating in-depth studies and the development of custom network configurations, O-RAN microservices, and experiments.
\end{itemize}

Table~\ref{tab:Table_combined} expands on the foundational concepts of an open-source SDR platform for cellular connected UAV experimentation introduced in our earlier research~\cite{itworks}. This updated version introduces the technical and operational requirements essential for effective O-RAN experimentation. Additionally, it has been thoroughly revised to incorporate the latest advancements in hardware technologies, which are critical for the successful deployment of a 5G standalone (SA) network. 

\subsection{
 Near-RT RIC Deployment Options for AERPAW}

AERPAW enables advanced wireless experiments with 3D mobility through UAVs. 
Its architecture is designed for experimentation through a single or multiple containers called the experimenter virtual machine (E-VM). Despite its name, the E-VM is a Docker container that allows for the deployment and testing of custom SDR setups and network configurations. An E-VM may implement a 5G user equipment (UE), a 5G base station or gNodeB (gNB), or the 5G core (5GC).  
Users get full access to the E-VM for programming or configuring an experiment; they can configure SDRs and UAV 
autopilots through APIs offered by standard programming libraries. Docker's containerization ensures that the E-VM is lightweight, portable between AERPAW's DT and physical testbed, and easily replicable across different testbed nodes, ensuring consistent and reproducible experiments~\cite{aerpawarch}.


Accessed through AERPAW's experimenter portal, users gain 
access to a range of resources such as software development tools and libraries, networking and radio software and hardware, and 
a DT~\cite{aerpaw_website}. 
By choosing available experimental software or developing custom experiments within E-VMs, 
one can iteratively test and refine the experiment on AERPAW's development environment, or DT, before transitioning the same E-VMs that are part of an experiment to the testbed’s operational environment~\cite{aerpawarch}.
The E-VM provides experimenters with root access, enabling control over SDRs and UAV/unmanned ground vehicle autopilots through APIs offered by standard programming libraries. Docker's containerization ensures that the E-VM is lightweight, portable between AERPAW's DT and physical testbed, and easily replicable across different testbed nodes, ensuring consistent and reproducible experiments.


The integration of O-RAN technologies and, specifically, the near-RT RIC and E2 interface currently \textcolor{black}{offer} two popular open-source options: \textcolor{black}{OpenAirInterface's (OAI's)} FlexRIC and \textcolor{black}{O-RAN Software Community's (OSC's)} RIC~\cite{NGO2024}. 
FlexRIC provides a simplified near-RT RIC setup that reduces deployment complexity, lowering the barrier for researchers new to O-RAN who are focused on rapid prototyping and innovation without the need for industry compliance. The FlexRIC architecture allows for high customization and direct control over experiments, accommodating varied research needs~\cite{10.1145/3485983.3494870}. The OSC RIC, on the other hand, is a robust and containerized RIC implementation that is compliant with the O-RAN Alliance industry standard. It, however, presents significant integration challenges due to its microservice-based deployment model. 

AERPAW’s reliance on Docker containers for testbed access, providing a streamlined, container-based experimentation framework where experimenters are offered  one E-VM per AERPAW node, complicates the integration of the OSC RIC, as its microservice architecture requires multiple containers and extensive orchestration
~\cite{AlyORAN}. The necessity for a substantial infrastructure setup to support OSC RIC's microservices can divert resources from primary research activities, making it less suitable for AERPAW's use case. In contrast, FlexRIC is inherently compatible with AERPAW's container-based architecture. It can operate efficiently within a single container, even when deployed alongside other essential components such as softwarized 5GC and access networks (Open5GS and OAI gNodeB in this case). This ability to consolidate multiple elements within a single container simplifies its deployment with the necessary interfaces into \textcolor{black}{AERPAW's software infrastructure, resource management, and experiment design flow}.

The FlexRIC integration presented in this paper enables seamless deployment of xApps for data collection or other monitoring or control functions within AERPAW's custom experiment lifecycle. 
This is the first integration of its kind in WCTs with a primary focus on UAV research. Table~\ref{tab:Table_combined} provides the technical specifications that we use for prototyping and testing. 

\begin{figure}[t]
\centering
\includegraphics[width=0.47\textwidth]{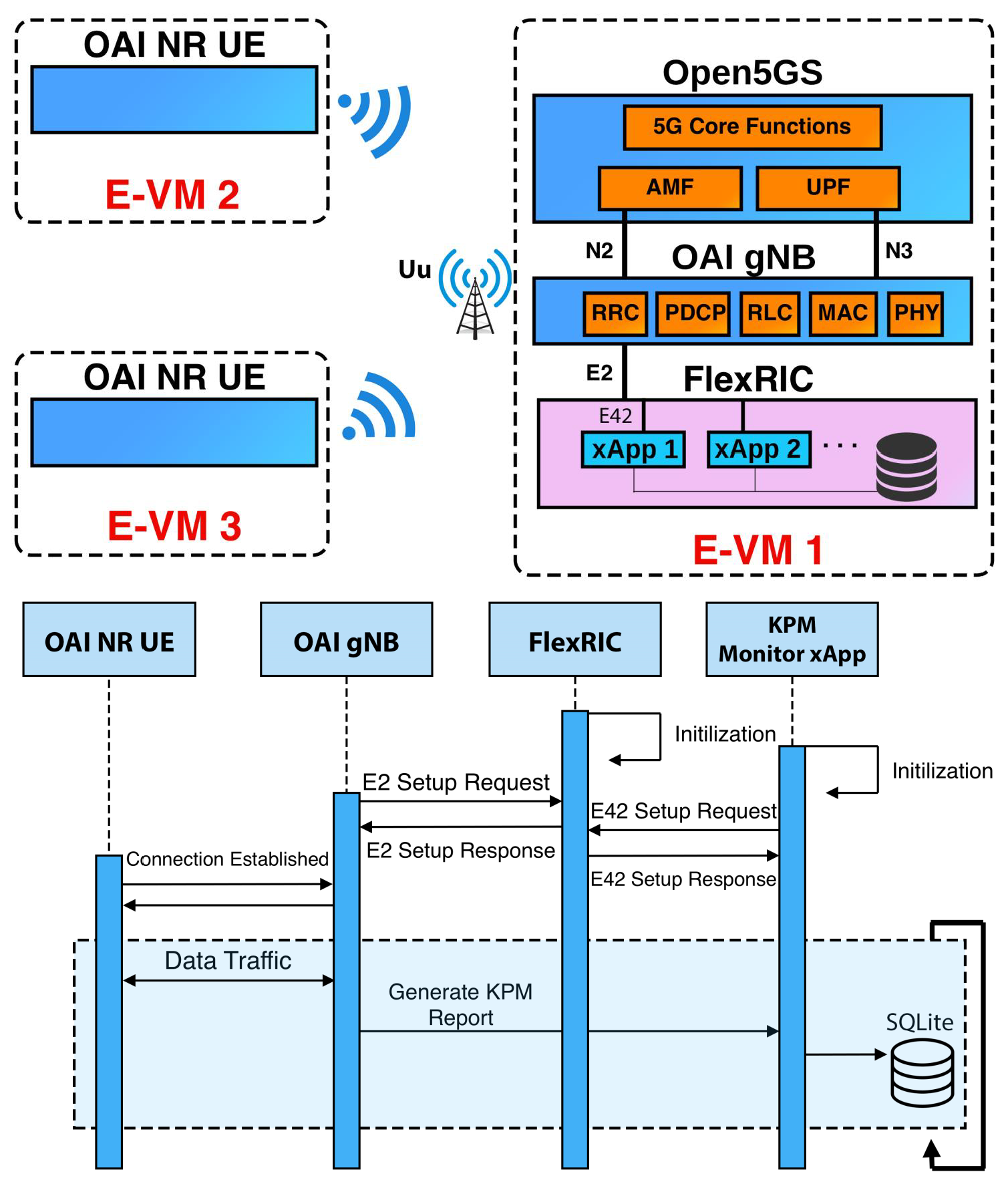}
\caption{System model (top) and sequence of events in the O-RAN reference architecture (bottom).}
\label{fig:setup}
\end{figure}

%% file: Tables/ReqTable.tex
\begin{table*}
\centering
\caption{Technical design requirements and  
specific components enabling 5G cellular networked UAV research with O-RAN.}
\footnotesize
\label{tab:Table_combined}
\resizebox{0.91\textwidth}{!}{
\begin{tabular}{|p{1.0cm}|p{1.6cm}|p{6.4cm}|p{7.5cm}|}
\hline
\multicolumn{2}{|c|}{\textbf{System Features}} & \textbf{Technical Design Requirements} & \textbf{Specific Components} \\
\hline
\textcolor{black}{Software-Defined Radio (SDR)} & Data Conversion &
 SDR hardware must support high-resolution \textcolor{black}{analog-to-digital and digital-to-analog converters (ADCs/DACs)} with sampling rates that can accommodate advanced modulation schemes across a broad RF spectrum. Driver compatibility and community adoption are essential for versatile deployment. &
B210 \textcolor{black}{Universal Software Radio Peripheral (USRP)}: Sampling Rate (max) 61.44 \textcolor{black}{mega-samples per second}, ADC and DAC Resolution 12 bits, 
Dimensions 9.7x15.5x1.5 cm, Weight 350 g. \\
\cline{2-4}
& Baseband/ Protocol Processing &
 Requires high-performance multi-core processors capable of handling intensive computational loads, equipped with high-speed data transfer interfaces. &
 Intel NUC13 (mobile node \& fixed node): 6-core i7-10710U, 32 GB RAM, Thunderbolt 3, USB 3.0. \\
\cline{2-4}
& Software &
 Support for advanced network management software, including both open and proprietary solutions, to facilitate a range of experimental configurations. &
 Configurations include USRP Hardware Driver v4.3.0, \textcolor{black}{OpenAirInterface (OAI)} latest branch, open5GS v2.6.4; all hosted on Dockerized Ubuntu 22.04 LTS. Linux   \\
\hline
RF Front End & Amplifier &
 Amplifiers ensure necessary transmit and receive power levels for the required coverage. Device choices and SDR gain settings need to minimize self-interference, signal distortion, heat generation and power consumption, particularly in mobile setups. &
 Power amplifier: Mini-Circuits (MC) ZHL-15W-422-S+ with 46 dB gain for fixed; MC ZVE-8GX+ with 30 dB gain for mobile node transmitters. 
 Low-noise amplifier: MC ZX60-83LN12+ with 22 dB gain for fixed and mobile node receivers.
\\
\cline{2-4}
& Filters &
 Must effectively suppress spurious emissions and prevent harmonic interference, configured specifically for the operational frequency bands. &
 MC VLF-4400+ (Tx): Low pass filter, DC-4400 MHz passband. MC VBFZ-3590+ (Rx): Bandpass filter, 3000-4300 MHz operational range. \\
\cline{2-4}
& Antennas &
 Require antennas that offer extensive frequency coverage and adaptability, with considerations for aerodynamic and weight constraints on UAV platforms. &
 Mobile Mark RM-WB1-DN-B1K for fixed; Octane Wireless SA-1400-5900 for mobile nodes providing robust coverage and minimal physical footprint. \\
\hline
Network & RAN, \textcolor{black}{core network (CN), and user equipment (UE)} & Must support a 5G network using open-source software 
implementing the gNodeB (gNB), 5G CN functions, and the UE. 
  & The 5G \textcolor{black}{standalone (SA)} deployment consists of Open5GS CN functions for managing authentication, mobility, and session control; the OAI gNB for handling radio access and resource allocation
; and the OAI New Radio UE, which enables user device connectivity and seamless interaction within the 5G network. \\
\cline{2-4}
& Experimental Network &
 Must support high-throughput, low-latency connections between RAN and CN, suitable for dynamic and distributed experimental environments. &
 Gigabit Ethernet or Wi-Fi 6 (AX4800) for wired or wireless backhaul ensuring robust, high-speed connectivity suitable for 
 scalable experiments with multiple gNBs. \\
\cline{2-4}
& Experiment Control &
 Secure and dependable control mechanisms are vital for managing UAVs and network elements remotely under varying experimental conditions. &
 Secured private \textcolor{black}{internet protocol-based} network utilizing \textcolor{black}{secure shell} over \textcolor{black}{Wi-Fi 6}, ensuring reliable and secure experiment control of mobile nodes. \\
\hline
O-RAN & O-RAN Interface &
 Must support O-RAN Alliance specified open interfaces like E2, A1, and O1/O2, ensuring integration with existing and future RICs and network elements. &
 OAI 5G SA deployment interfaces with the FlexRIC \textcolor{black}{near-RT RIC} using an O-RAN complaint E2 interface. \\
\cline{2-4}
& O-RAN RICs &
 Must provide capability for RAN optimizations leveraging O-RAN RICs, supporting AI-driven network management. &
 FlexRIC supports a \textcolor{black}{near-RT RIC} terminated by the E2 interface and capable of hosting iApps and xApps. FlexRIC is also capable of integrating with an O-RAN Software Community RIC.   \\
\cline{2-4}
&  RAN Intelligent Control &
 RIC apps such as xApps and rApps should enable specific optimization and control. &
 \textcolor{black}{Key performance indicator monitoring} xApp, which provides comprehensive performance metrics for the RAN. Additionally, we utilize a monitoring xApp that gathers metrics from the \textcolor{black}{medium access control, radio link control, and packet data convergence protocol} layers, leveraging the \textcolor{black}{key performance measurement service model} for enhanced performance monitoring.\\
\cline{2-4}
& O-RAN \textcolor{black}{Functional Split} &
 Must provide functional split capabilities that allow for the separation of radio functions between the DU and CU to meet specific deployment needs.  &
 OAI project supports multiple functional splits, including 7.2, 6.2, and 2. We use a traditional deployment of the \textcolor{black}{gNB} to support AERPAW's single container \textcolor{black}{experimenter virtual machine}.\\
\cline{2-4}

\hline
\end{tabular}}
\end{table*}

%% file: include/data.tex
\begin{figure}[t]
\centering
\includegraphics[width=0.39\textwidth]{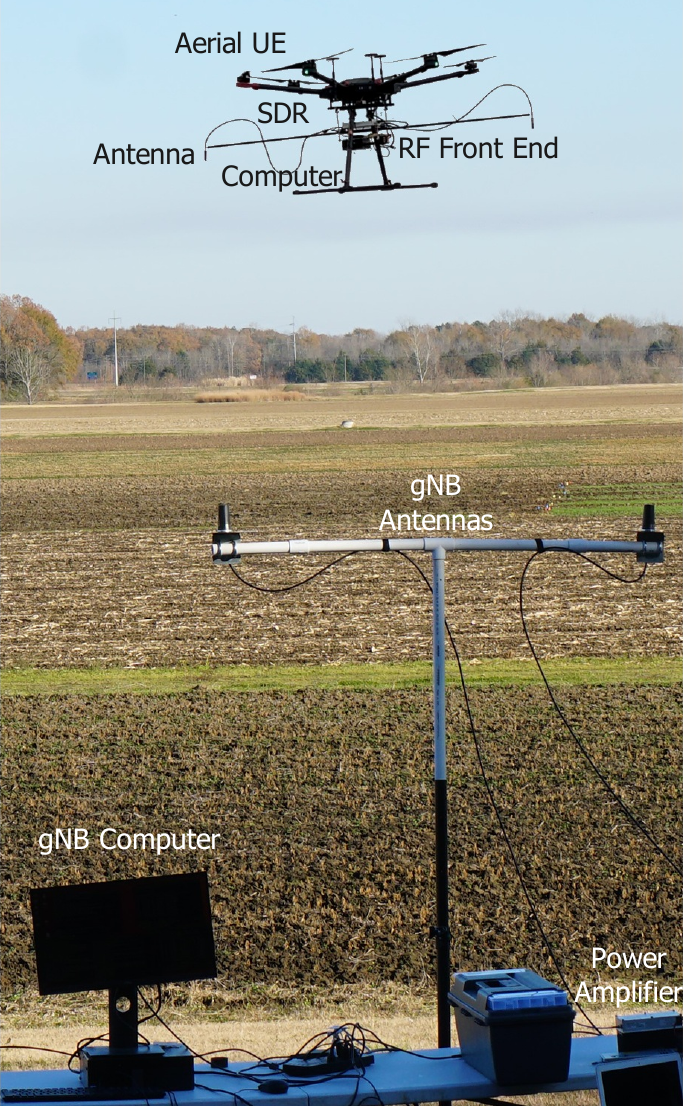}
\caption{Experimental deployment showing the gNB and AUE. 
}
\label{fig:5gota}
\end{figure}


Our O-RAN software deployment as captured by Fig.~\ref{fig:setup} details how FlexRIC is to be integrated into the AERPAW testbed and DT and used to collect data from a 5G RAN, or to control the RAN. 
In our deployment, E-VM~1 hosts the 5GC (Open5GS) 
paired with the OAI 5G gNB. This configuration serves as the core 5G infrastructure of AERPAW, providing essential connectivity and network services to both UAV and ground nodes. FlexRIC, integrated in the same E-VM, connects through the E2 interface to the E2 node (gNB). It hosts xApps and a database which enables storing and retrieving data collected and used by different xApps. E-VM~2 and E-VM~3 \textcolor{black}{both} host the OAI New Radio UE and are deployed on 
a UAV and a fixed ground node, respectively. 
The sequence of events can be seen at the bottom of Fig.~\ref{fig:setup}, which details how data is collected from the E2 node and stored in an SQLite database for offline processing and analysis or for other xApps to use. 


The E2 interface connects FlexRIC to the gNB, enabling the exchange of both performance metrics and control commands. xApps deployed on \textcolor{black}{FlexRIC subscribe to specific service models, such as the key performance measurement service model}, via the E2 Application Protocol. 
Shown in our experiments is one use case which is the collection of critical metrics including throughput and latency. Additional metrics, such as \textcolor{black}{Global Positioning System (GPS)} tracking provide 3D coordinates of an aerial UE (AUE). These metrics are gathered periodically by the RAN or triggered by specific network events, and \textcolor{black}{are} then sent to FlexRIC for storage in its internal database. The E2 interface supports near-instantaneous 
communication, enabling continuous, near-RT 
network monitoring and control. Internally, FlexRIC communicates with xApps through a lightweight interface (E42), ensuring efficient data exchange. This architecture provides reliable logging and retrieval of collected data, facilitating post-experiment analysis and empowering xApps to drive network, service, or UAV trajectory optimizations, to name a few, enhancing the responsiveness and performance of the network or particular service.


Our experimental 
setup involves two UEs: one ground node at approximately 20 m distance from the gNB at 1 m height and a mobile AUE. 
The gNB antennas are at 2.5~m height above ground on an slightly elevated area as shown in Fig.~\ref{fig:5gota}. The UAV follows a predefined flight path. 

For the gNB, we use a \textcolor{black}{time division duplex} configuration with a 30 kHz subcarrier spacing and a 5 ms downlink/uplink (DL/UL) periodicity. In each 5 ms period, 7 full DL slots and 6 additional DL symbols are allocated for DL transmission, totaling 104 DL symbols, while 2 full UL slots and 4 additional UL symbols are allocated for UL transmission, totaling 32 UL symbols. This configuration results in an approximate UL/DL ratio of 0.31. 
The center frequency is 3.5 GHz in \textcolor{black}{5G New Radio Band n78} with all measurements taken on the DL. Without loss of generality, the target DL throughputs are 2 and 18 Mbps for the ground UE and AUE, respectively. 

\subsection{Average Data Rate Versus Distance}

\begin{figure}[t]
\centering
\includegraphics[width=0.47\textwidth]{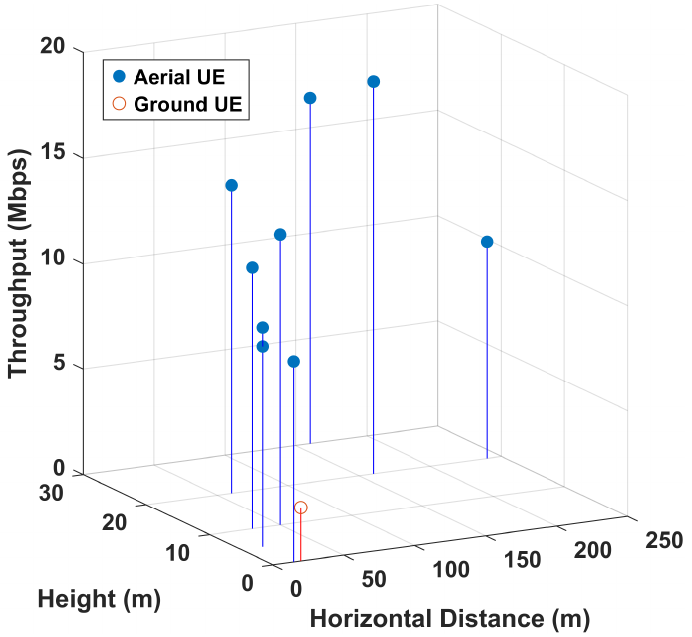}
\caption{Average data rates for a fixed ground UE and AUE at selected UAV hovering positions. The ground base station is at the origin. 
}
\label{fig:avgdatavsdistance}
\end{figure}

From Fig.~\ref{fig:avgdatavsdistance}, it is evident that the AUE node maintains a consistent data transmission rate of around 10 Mbps when positioned at distances of 15 and 20 m away from the base station while maintaining a fixed altitude of 5 m. This achieved data rate starts to increase and reaches approximately 13 Mbps as the AUE elevates its altitude from 5 to 10 m, despite the horizontal distance extending to 30 and 50 m from the gNB, which is located at the origin in Fig.~\ref{fig:avgdatavsdistance}. This performance trend remains applicable as the AUE continues to increase its altitude from 10 to 20 m, even as its distance to the gNB increases. 
This performance can be attributed primarily to the irregular 3D antenna pattern and lack of antenna diversity in this setup. In general, the line of sight (LoS) probability improves which increasing AUE altitude. Despite the strong LoS, 
the path loss increases with distance, leading to a decrease in the achievable data rate. 
As shown in Fig.~\ref{fig:avgdatavsdistance} the measured data rate of the gNB-AUE DL fluctuates between a minimum of 5 Mbps and the 18 Mbps target as a result of the varying channel conditions. 

\subsection{Resource Block Allocation}
RBs are allocated dynamically in OAI based on a scheduler that optimizes the balance between fairness and throughput. The scheduler uses channel quality indicator reports to determine the \textcolor{black}{modulation and coding scheme}. 
Based on these estimates, it calculates a proportional fairness (PF) coefficient, which reflects the potential bitrate each UE can achieve in a given subframe. The PF coefficient is adjusted according to the amount of channel usage by each UE, with priority given to those that have used the channel less recently. The scheduler allocates RBs to the UE with the highest PF coefficient until resources are exhausted or the UE no longer requires additional RBs. The process then continues with the next UE having the highest PF coefficient~\cite{oaischeduler}. 

\subsection{Service Data Unit Latency}
\begin{figure}[ht]
\centering
\includegraphics[width=0.48\textwidth]{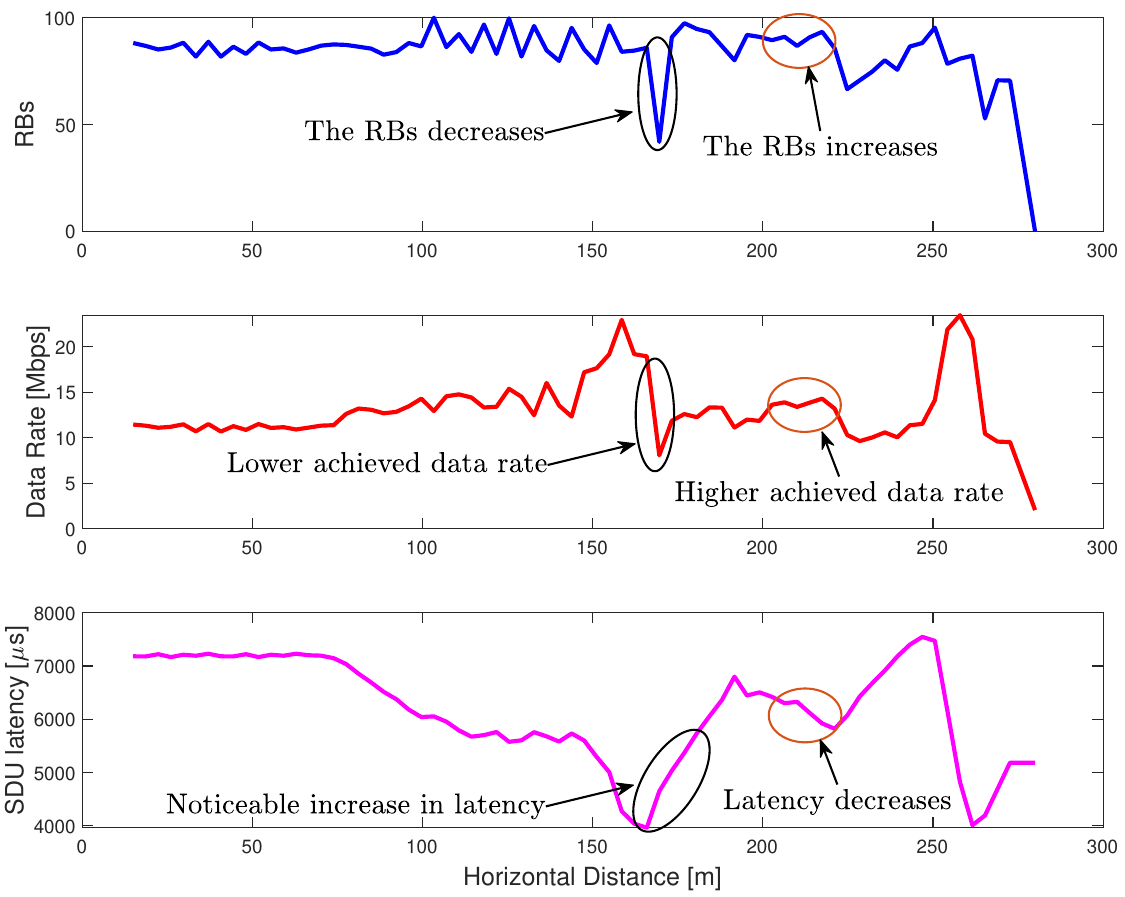}
\caption{RB allocation, achieved DL data rate, and SDU latency over horizontal distance to the gNB for a AUE flying at a continuous speed of 10~m/s (no ground UE in this experiment).}
\label{fig:LatencyratevsRBs}
\end{figure}

In the FlexRIC framework, an additional metric quantifies the delay that the service data units (SDUs) experience in the DL at the \textcolor{black}{radio link control (RLC) layer}. This measurement is vital for evaluating how efficiently the RLC layer handles data, impacting overall network performance, especially for maintaining quality of service in 5G networks. 
Fig.~\ref{fig:LatencyratevsRBs} shows the RB allocation, DL data rate, and SDU latency figures for the AUE over the course of the UAV flight. 
Within the initial 150 m horizontal distance to the gNB, an inverse relationship is observed between data rate and SDU latency: As the data rate increases for a given offered load, packets wait less time in the transmit buffers. Once the 18 Mbps throughput target is achieved (160 and 260 m), the scheduler reduces the number of RBs allocated to the AUE, which subsequently leads to a data rate drop, increasing SDU latency. 

\subsection{Advanced DT for Data Generation Leveraging O-RAN}

\begin{figure}[t]
\centering
\includegraphics[width=0.47\textwidth]{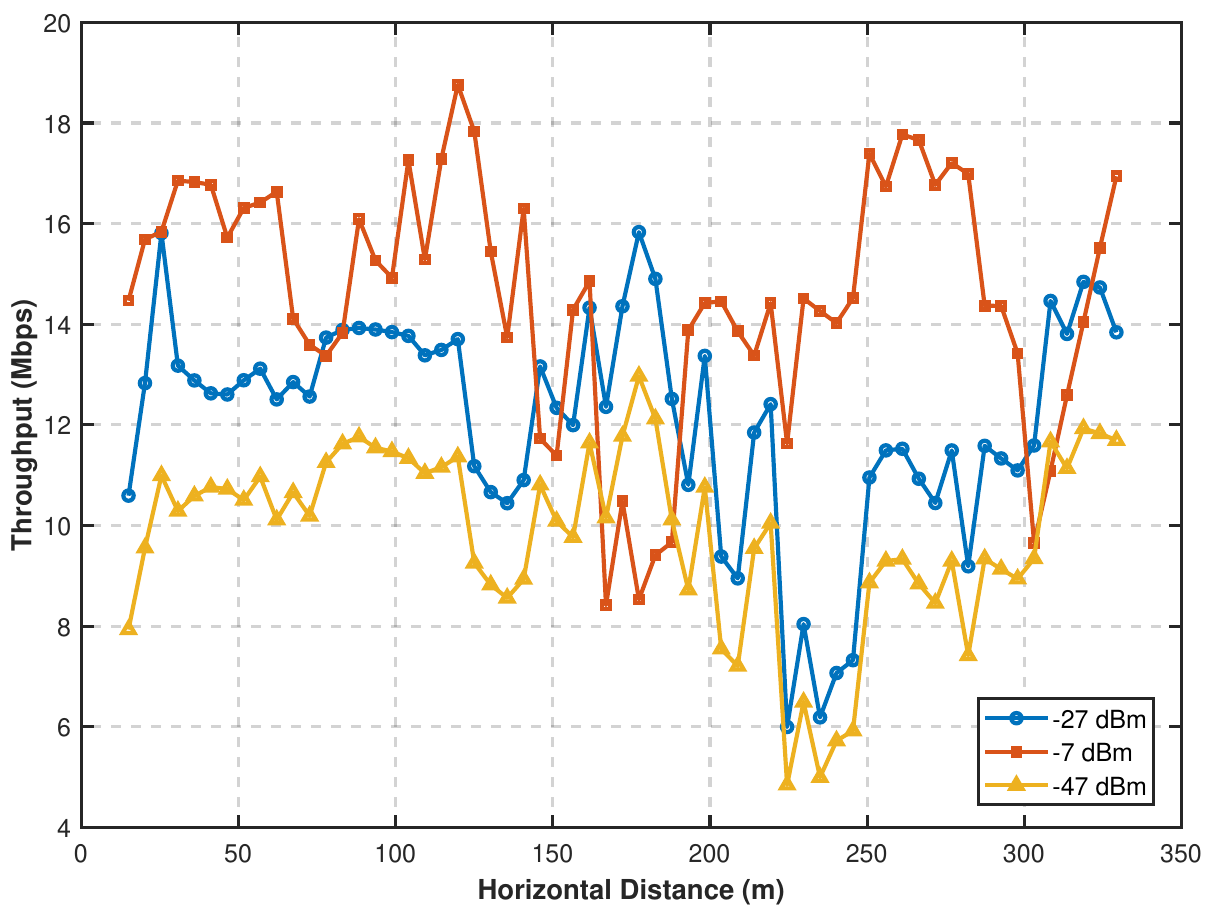}
\caption{Measured (black) and generated (red and yellow) throughput over horizontal distance between the AUE and the gNB for different maximum PDSCH reference signal power parameter values set in the OAI configuration for the B210 \textcolor{black}{Universal Software Radio Peripheral}.} 

\label{fig:gpt}
\end{figure}

While DTs have been extensively studied for wireless networks, their integration with autonomous vehicles with programmable mobility, such as UAVs, is largely unexplored. 
Here we showcase the potential of 
GAI to create datasets 
that may be used to aid a DT in terms of data generation for various scenarios rather than validating a specific DT. The use of AI to generate data aligns with recent literature advocating for AI integration within AERPAW's DT, enhancing emulation accuracy and experiment diversity, particularly in dynamic environments such as autonomous vehicle networks
~\cite{gürses2024digital}. 
The GAI model employed in this study is ChatGPT 4, which is used to simulate varying gNB transmit power levels. 
Using few shot prompting, pre-trained models can be highly accurate in producing \textcolor{black}{a} high-quality output to meet the objectives of the given instruction \cite{lee2024llmempoweredresourceallocationwireless}. We showcase results derived from a real-world dataset capturing the relationship between throughput and distance for an AUE, while leveraging GAI to simulate varying transmit power levels. 
We provide real data collected for a single power level, -27 dBm, and prompt the model to generate throughput (Mbps) at the same distances for various other power levels. 
We consider other open and closed source GAI models, each with parameter counts in the billions, and find that ChatGPT 4 achieved a score of 100\% with 1760 parameters, followed by Chat GPT 3.5 with a score of 95\% and 154 parameters. Sus-Chat also scored 95\%, but with only 34 parameters, while both Llama and Mistral scored 70\%, each having 7 parameters. 

%% file: include/challenges.tex
In the development and deployment of a 5G O-RAN-based testbed for UAV research, several challenges and research directions emerge, each requiring careful consideration and innovative solutions. 

\subsection{O-RAN Integration Challenges}
\begin{itemize}
    \item \textbf{Interoperability and Standardization Challenges:} Achieving seamless and efficient interaction among diverse network components from different vendors 
    is challenging. Developing robust testing frameworks that can assess component compatibility under varying conditions is critical. Research should focus on creating automated testing tools to preemptively resolve such issues.
    \item \textbf{RT Data Processing and AI Integration:} UAV operations require dynamic and RT decision-making. Integrating AI-driven solutions for network management and data processing presents significant challenges. Efficient data processing architectures and AI algorithms need development to operate at the network edge, enhancing responsiveness and ensuring data privacy and security.
    \item \textbf{Scalability Challenges:} 
    AERPAW's single-container approach limits the system's scalability and flexibility, particularly 
    since the O-RAN architecture is designed on the principles of microservice based deployments. In single-container environments, resource isolation and parallelism are constrained making it difficult to effectively distribute workloads. 
    The inability to scale the architecture dynamically limits experimentation with high-throughput or low-latency services, which are critical in 5G and UAV network deployments.
    A multi-container E-VM setup, where key components---FlexRIC, RAN, and \textcolor{black}{core network}---are decoupled into separate containers, would allow for better isolation, resource management, and independent scaling of O-RAN functions. However, moving to a multi-container setup introduces new complexities, such as ensuring low-latency inter-container communication, managing synchronization between RAN components, and maintaining consistency across the control and data planes. Achieving this requires implementing a robust orchestration framework, such as Kubernetes, as well as employing container networking strategies such as service meshes to efficiently handle inter-container data flows. 

\end{itemize}

\subsection{O-RAN R\&D Opportunities} 
O-RAN facilitates 
advancing UAV integration into cellular networks and the deployment of tailored services, with critical R\&D needed in the following areas. 
\begin{itemize}
    \item \textbf{Diverse UAV Application Support:} Research on dynamic network slicing 
    can ensure that UAVs operate under optimal conditions for specific applications such as surveillance. 
    This involves developing slice lifecycle management tailored for aerial scenarios.
    
    \item \textbf{Secure UAV Communications:} An important research direction involves the development of secure communication protocols tailored to UAV constraints, such as limited processing power, to ensure data integrity and security in sensitive environments. These protocols need to withstand attacks and unauthorized access, providing a secure communication framework for UAV operations. 
    \item \textbf{Spectrum Management:} Efficient spectrum management and interference mitigation techniques are critical to prevent signal interference between aerial and terrestrial network users, 
    ensuring reliable communications and services. 
    Research needs to explore 
    spectrum sharing 
    techniques, leveraging O-RAN's microservices, to navigate and contribute to the evolving regulatory landscapes and maximize network/spectrum efficiency.
\end{itemize}

%% file: include/conclusions.tex
This article has evaluated open source near-RT RIC options for their integration into AERPAW and proposed the integration of FlexRIC into the testbed and DT. 
FlexRIC, the E2 interface, and a key performance indicator monitoring xApp have been deployed in AERPAW's E-VM. 
We have shown initial results from various air-to-ground 5G experiments in a real world setting and discussed research challenges and opportunities with the goal of enabling data-driven UAV network optimization R\&D. Our prototyping reveals that O-RAN experimentation capabilities can be integrated into AERPAW to support 6G UAV communications research experiments related to open and AI-enhanced RAN architectures.